# Synthesis of Polar Ordered Oxynitride Perovskite

**Rajasekarakumar Vadapoo[1], Muhtar Ahart[1], Maddury Somayazulu[1], Nicholas Holtgrewe[1,2], Yue Meng[3], Zuzana Konopkova[4], Russell J. Hemley[5], and R. E. Cohen[1,6,*]**

[1]Extreme Materials Initiative, Geophysical Laboratory, Carnegie Institution of Washington, Washington, DC 20015, USA

[2]Department of Mathematics, Howard University, Washington DC 20059, USA

[3]High Pressure Collaborative Access Team, Geophysical Laboratory, Carnegie Institution of Washington, Argonne, Illinois 60439, USA

[4]Photon Science DESY D-22607 Hamburg, Germany

[5]Department of Civil and Environment Engineering, The George Washington University, Washington DC 20052, USA

[6]Department of Earth and Environmental Sciences, Ludwig Maximilians University, Munich, Germany

[*]rcohen@carnegiescience.edu

**ABSTRACT**

For decades, numerous attempts have been made to produce polar oxynitride perovskites, where some of the oxygen are replaced by nitrogen, but a polar ordered oxynitride has never been demonstrated. Caracas and Cohen studied possible ordered polar oxynitrides within density functional theory (DFT) and found a few candidates that were predicted to be insulating and at least metastable. $YSiO_2N$ stood out with huge predicted polarization and nonlinear optic coefficients. In this study, we demonstrate the synthesis of perovskite-structured $YSiO_2N$ by using a combination of a diamond anvil cell and in-situ laser heating technique. Subsequent in-situ X-ray diffraction, second harmonic generation, and Raman scattering measurements confirm that it is polar and a strong nonlinear optical material, with structure and properties similar to those predicted by DFT.

**Introduction**

Oxide perovskites are common and form with cations from a majority of the elements of the periodic table, and polar perovskites form extremely interesting and useful active materials, with applications that include nonlinear optics, ferroelectric memories, piezoelectric transducers and actuators. For decades, numerous attempts have been made to produce polar oxynitride perovskites, where some of the oxygen are replaced by nitrogen, but a polar ordered oxynitride has never been demonstrated. (repeated from abstract) Oxynitrides have been extensively studied for their improved electronic properties. [1, 2, 3, 4, 5] Among the wide variety of synthesised oxynitrides, $CaTaO_2N$ and $LaTaON_2$ have been proposed for nontoxic pigments [6] and $BaTaO_2N$ as a high dielectric [7] and catalyst for photoelectrolysis of water. [8] $EuNbO_2N$ and $EuWON_2$ show intrinsic colossal magnetoresistance. [9,10] Anion ordering could produce polar oxynitrides, but making such materials in the lab had been unsuccessful until now. $SrTaO_2N$ bulk [7] does show high dielectric properties, and epitaxial thin films have been reported to have possible ferroelectricity. However, whether it is intrinsic is still unclear. [11] Recently, $MnTaO_2N$ was reported to have a polar structure with helical spin ordering without anion-ordering. [12]

Anion ordering in insulating oxynitrides would provide polar, electrically active materials. [3 13] Thus considerable efforts have been exerted to synthesise oxynitrides with ordered anions. [10, 14, 15, 16, 17, 18, 19] Oxynitrides are usually synthesised using ammonolysis, [20, 21, 22, 23] which generally yields nonordered, nonpolar, centrosymmetric materials. Caracas and Cohen predicted a new class of stable polar oxynitride perovskites of ordered structure, including yttrium silicon oxynitride, $YSiO_2N$, [3, 5] using a materials-by-design approach. They searched through many nontransition element-bearing $ABO_2N$ compositions using density functional theory (DFT) and considered those that were insulating and stable via energy minimisation in a possibly distorted perovskite structure.

Phonons were then computed using density functional perturbation theory (DFPT) for those that met these abovementioned criteria. Among the few that had stable phonon modes, YSiO$_2$N demonstrated superior properties. It was predicted to be stable in a simple 5 atom cell polar structure with symmetry P4mm (Fig. 1). DFPT also predicted that YSiO$_2$N has a giant effective spontaneous polarisation of 130 mC/cm$^2$ and a very large nonlinear optic coefficient.

**Experiments**

The high pressure-temperature experiments for the synthesis and in-situ and ex-situ X-ray diffraction measurements were conducted at the High Pressure Collaborative Access Team's monochromatic angle-dispersive X-ray diffraction station 16-ID-B, sector 16 of APS, Argonne, Illinois, United States, and at the light source Petra III at DESY, Hamburg, Germany. Finely ground yttrium nitride (YN) and amorphous silica powders were loaded into a Diamond Anvil Cell DAC with a steel gasket and ruby as the pressure standard. Simultaneous high *P-T* conditions were achieved using the online laser-heated DAC system.[24] Temperatures were determined by fitting the Planck radiation function to the thermal radiation collected from the heated sample. For ex-situ X-ray diffraction measurements, a 30 KeV X-ray beam was focused to 7 to 5 microns in horizontal and vertical directions, and diffraction patterns were recorded with a MAR charge-coupled device (CCD) detector or on a Perkin-Elmer detector. Details on the laser heating system and sample preparations can be found in the Supplemental Information.[25]

Raman spectroscopy studies were carried out using a 532 nm laser as the excitation source in a backscattering geometry. The sample in the DAC was placed in a microscopic stage in our Raman setup: a Mitutoyo 20X objective with a numerical aperture of 0.28. Our microscopic Raman system, with a spatial resolution of 1.14 mm, was used for both incident light focusing and collection of scattered light; subsequently, scattered light was sent to a monochromatic spectrometer (Princeton Instrument: Acton SP2300 spectrometer) via a confocal optical pass, and spectra were recorded with a CCD detector.

For the SHG measurements near IR (1,064 nm, Nd: YAG), 8 ns to 20 ns pulsed laser with a 1 kHz to 20 kHz repetition rate was used to excite the SHG signal; a dedicated spectrograph equipped with a CCD detector synchronised with the laser was used for the SHG signal acquisitions. The acquisition time was approximately 1 s. All measurements were performed at room temperature. More details are given in the Supplemental Information. [25]

## Results

**Synthesis and structural characterisations**

We used the synthesis method outlined by Caracas and Cohen, using YN and $SiO_2$ as reactants: YN + $SiO_2$ ($YSiO_2N$). [3,5] We finely ground parent yttrium nitride (YN) and amorphous silica ($SiO_2$) in an argon atmosphere, which is necessary because of the reactive nature of YN with air and moisture. We loaded the powders in a diamond anvil cell (DAC) in a stack geometry of $SiO_2$/YN/$SiO_2$ to enhance the diffusion process of starting materials. The sample was compressed to high pressure and then heated using the double-sided laser heating method. [24] The resultant sample with the X-ray spot size of 4 μm was then studied in-situ and ex-situ using optical microscopy as well as synchrotron X-ray diffraction at the Advanced Photon Source (APS) beam line 16-ID-B and at the light source Petra III at DESY, Hamburg, Germany. A new phase formed at pressures as low as 4 GPa and above 1,500 K along with other minor phases. We analysed the reaction products using synchrotron X-ray diffraction and micro-Raman spectroscopy.

The sample of YSiO$_2$N was synthesised by reacting a mixture of cubic YN and SiO$_2$ glass. From our earlier in-situ laser heating experiments conducted at HPCAT of APS, we realised that an excess of one of the two components is preferable to laser heating a 1:1 mixture. Second, we wanted to minimise the interference of a more complex coesite (monoclinic) in comparison to a tetragonal stishovite phase in the diffraction pattern and accordingly choose a synthesis pressure higher than 10 GPa. Since the current experiment involved ex-situ synthesis, we decided to obtain diffraction patterns from a grid encompassing the whole sample chamber both before and after the heating cycle. The laser heating was performed with a 1.06 mm Nd: YLF fibre laser focused to a 30 μm heating spot. To minimise the effects of performing ex-situ synthesis, we rastered the whole sample chamber with the heating beam. Although the initial pressure was estimated to be 13 GPa, the pressure after synthesis dropped to 10 GPa, which could be attributed to release of stresses at high temperatures. Diffraction pattern from the quenched sample showed regions of unreacted YN as well as a highly crystalline and textured pattern of YSiO$_2$N. Stishovite and minor amounts of coesite were also observed with much less texture since they were synthesised from a micro-grained, glassy starting material. This is shown by the cake pattern (Fig. 2a). The predominant phases are labelled accordingly in this 2D pattern as well as in the diffraction pattern (Fig. 2b) obtained from an integration of some of the azimuthal scans in this pattern. The starting material YN is highly granular (about 20 μm); the pattern becomes more granular after laser heating as well as highly textured. Given this innate poor quality of the diffraction patterns both for the starting material and the synthesized material, we found that any attempt to refine the patterns to extract structural information is fraught with difficulties and likely to give us a wrong assessment (whether about compositional disorder or the structure it-self). In fact, we don't see the strong [111] peak in the integrated powder pattern, but we do observe a few strong spots in the 2D image indicating that the synthesized sample is probably oriented preferentially along (111) direction.

That said, our diffraction data is consistent with the synthesised polar structure of YSiO$_2$N (space group P4mm), with lattice parameters a = 3.234(5) Å and c = 4.339(5) Å (Table 1), which match well with the theoretical prediction and proves the formation of predicted ordered YSiO$_2$N structure. [3,5] The high c/a ratio of 1.34 is strong evidence for O/N ordering. Theory shows that disordered YSiO$_2$N would likely be cubic perovskite (or transform to another phase), since it is the strong N-Si covalent bond that causes the distortion. Partially ordered YSiO$_2$N would have c/a ratios between 1 and 1.34. The Raman frequencies would also change significantly for disordered O and N.

From other runs, we synthesised this new polar phase of YSiO$_2$N at 12 GPa and 1,200 K. [26,27] Raman spectra of the above synthesised material measured at 3 GPa at ambient temperature is consistent with the theoretically predicted Raman spectra (Fig. 3 [a]). Most of the major Raman modes clearly match with the theoretically predicted modes represented by blue lines. [3,5] The predominant line observed 246 cm$^{-1}$ is in good agreement with the predicted mode whereas some modes show a small shift (Table 2). We also observed a strong peak at 105 cm$^{-1}$ and emergent peaks such as at 161 cm$^{-1}$ and 356 cm$^{-1}$ that are assigned to the coesite. [26] The agreement of the observed major Raman modes with the theoretically predicted ones further prove the formation of the oxynitride perovskite of YSiO$_2$N.

**Second Harmonic Generation**

Conventional X-ray or Raman cannot prove lack of inversion symmetry. We tested the synthesised material for inversion symmetry using optical second harmonic generation (SHG). We observed a very strong SHG signal (Fig. 3 [b]), clearly demonstrating that the synthesised material was indeed polar. For a crystal of thickness ($l$), the second harmonic intensity ($I$) is given by $I \propto I_{in}^2 f(n) d_{eff}^2 \Delta l^2 sin^2(\frac{\pi l}{2\Delta l})$, where $I_{in}$, $f(n)$, $d_{eff}$, and $\Delta l$ are the incident light intensity, the function of refraction indices, the effective nonlinear coefficient, and the coherence length, respectively.[28] Calculating the nonlinear optical coefficients could be a cumbersome process because of the very small size of the sample and the dependency of $f(n)$ and $d_{eff}$ with pressure. However, it should be noted that the SHG intensity observed in YSiO$_2$N at 11 GPa is comparable with our earlier studies on PbTiO$_3$ at 5 GPa for the similar incident signal. The synthesised YSiO$_2$N shows robustness of polarity against the pressure.

## Conclusions

We have successfully synthesised the unique class of polar ordered perovskite oxynitride predicted by the systematic chemical approach of the first-principles method. The polar ordered YSiO$_2$N has been achieved by high pressure-temperature methods. Our methodology and results provide an alternative way to synthesize new-type of materials which can be predicted with computational calculation but cannot be synthesized by traditional high-temperature solid-state reaction methods. Our results are consistent with synthesis of an anion ordered perovskite Oxynitride, possibly useful for non-linear optic applications.

## Acknowledgements

This work was designed and developed under the Office of Naval Research Grant No. N00014-14-1-0561. The instrumentations employed in the experiments, including the X-ray diffraction performed at HPCAT (Sector 16), Advanced Photon Source (APS), Argonne National Laboratory, were supported by the Carnegie/ Department of Energy Alliance Centre (CDAC: DENA0002006). HPCAT operations are supported by DOE-NNSA under Award No. DE-NA0001974 and DOE-BES under Award No. DE-FG02-99ER45775, with partial instrumentation funding by National Science Foundation. The APS is a U.S. Department of Energy (DOE) Office of Science User Facility operated for the DOE Office of Science by Argonne National Laboratory under Contract No. DE-AC02-06CH11357. Portions of this research were carried out at the light source PETRA III at DESY, a member of the Helmoltz association (HGF). We acknowledge support from the Carnegie Institution and the European Research Council advanced grant ToMCaT. We thank Hiroyuki Takenaka for helpful discussions.

# Table captions

**Table 1.** Lattice parameter observed for the synthesised material at 13 GPa after quenching from above 1,500 K. Theoretically predicted cell parameters of $YSiO_2N$, [3,5] and experimentally reported cell parameter of stishovite at 9.6 GPa. [29]

| Material | Lattice parameter (Å) | | | |
|---|---|---|---|---|
| | Observed @ 10 GPa | | Previous studies | |
| | a | c | A | c |
| Polar $YSiO_2N$ | 3.234(5) | 4.339(5) | 3.228 | 4.435 |
| Stishovite | 4.128(1) | 2.510(2) | 4.129 | 2.649 |

**Table 2.** Raman modes observed for the synthesised polar $YSiO_2N$ compared with the theoretically predicted modes along with their symmetry. [3,5]

| Experiment | Theory[3,4] | |
|---|---|---|
| | Symmetry | Modes (cm$^{-1}$) |
| 357 | A1(z) | 373 |
| | | 415 |
| 652 | | 648 |
| | | 750,927,1058 |
| | B1 | 400 |
| 246 | E(x,y) | 249 |
| 285 | | 281 |
| 381 | | 380 |
| 402 | | 402 |
| | | 534,721,854,930 |

# Figure captions

**Fig. 1.** P4mm polar structure of the ordered oxynitride perovskites ($YSiO_2N$) predicted by Caracas and Cohen.[3] Blue spheres represent yttrium (Y), green spheres inside the octahedra represent silicon (Si) and red and gold spheres represent oxygen and nitrogen, respectively.

**Fig. 2.** X-ray diffraction image and pattern (X-ray wavelength 0.4838 Å) of a sample synthesised at 13 GPa and well above 1,500 K and then quenched to ambient temperature and measured at 10 GPa. (a) Image of the diffraction lines (white vertical lines) with corresponding prominent indexed diffraction peaks marked with colour vertical bars for $YSiO_2N$ (red), stishovite (black) and yttrium nitride (blue). Along with these major contributions of the three phases, trace amounts of hexagonal $YSiO_2N$ and coesite were also detected and added to the phases fitted in the diffraction analysis. (b) In the pattern, marks represent the diffraction data, and the solid fitting curve represents the total simulated pattern using Le Bail fitting. The vertical bars from bottom to top show the corresponding diffractions for the new ordered phase of $YSiO_2N$ (black vertical lines), stishovite, which is the high-pressure phase of $SiO_2$ (red vertical lines), unreacted yttrium nitride (blue vertical lines), the high-pressure phase of $SiO_2$ coesite (green vertical lines) and hexagonal $YSiO_2N$ (brown vertical lines).[30,31] The difference between the observed and simulated pattern is shown in the lower part near the horizontal axis. The synthesised structure of polar $YSiO_2N$ has the structural parameters of the P4mm structure as predicted.[3]

**Fig. 3.** The new phase of $YSiO_2N$ synthesised at 12 GPa and 1,200 K. (a) Raman spectra measured at 3 GPa pressure at ambient temperature. The vertical bars from top to bottom show the observed experimental modes (red vertical bars) of the polar structure and the corresponding theoretically predicted Raman modes (blue vertical bars).[3] The brown star symbols correspond to the coesite,[26] the high-pressure phase of $SiO_2$ formed along with the synthesised mixture. (b) Observation of strong optical second harmonic generation measured at 11 GPa and ambient temperature, excited by a near IR 1,064 nm pulsed laser.

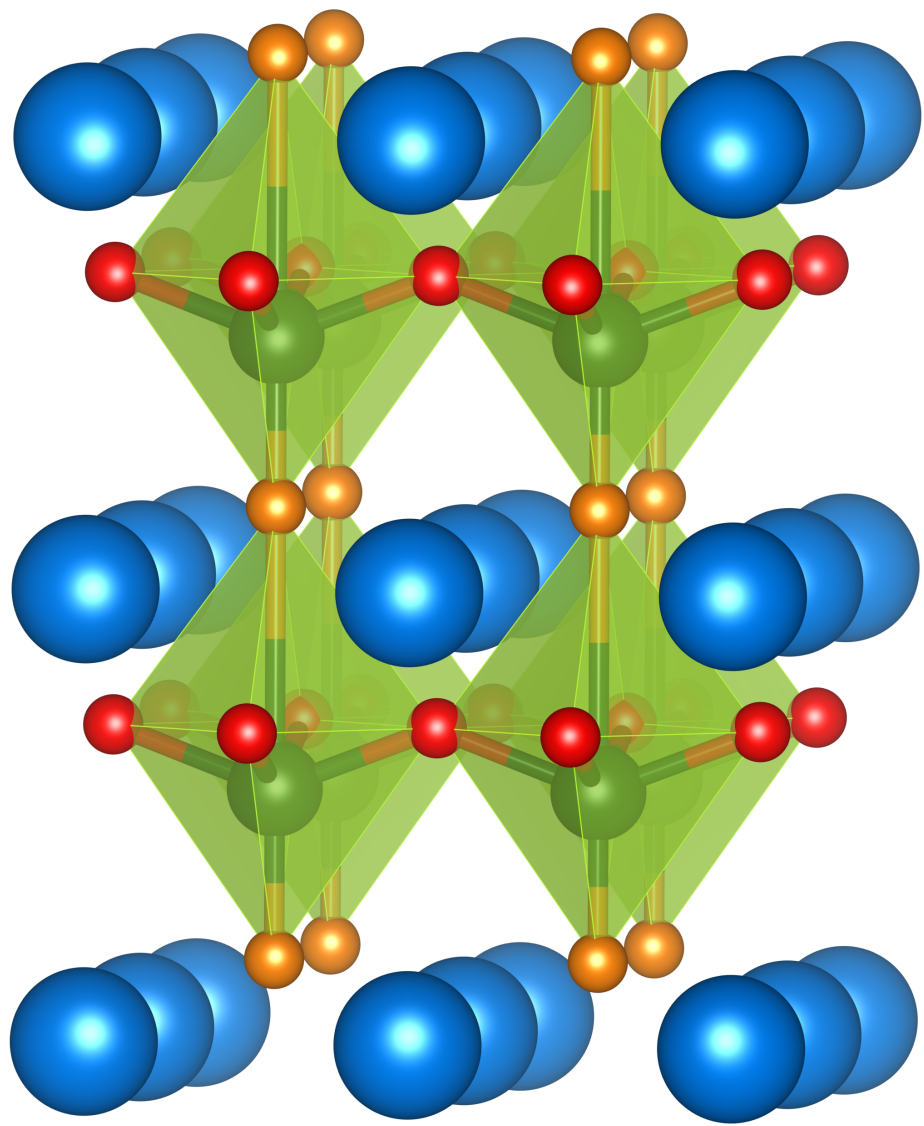

Fig. 1

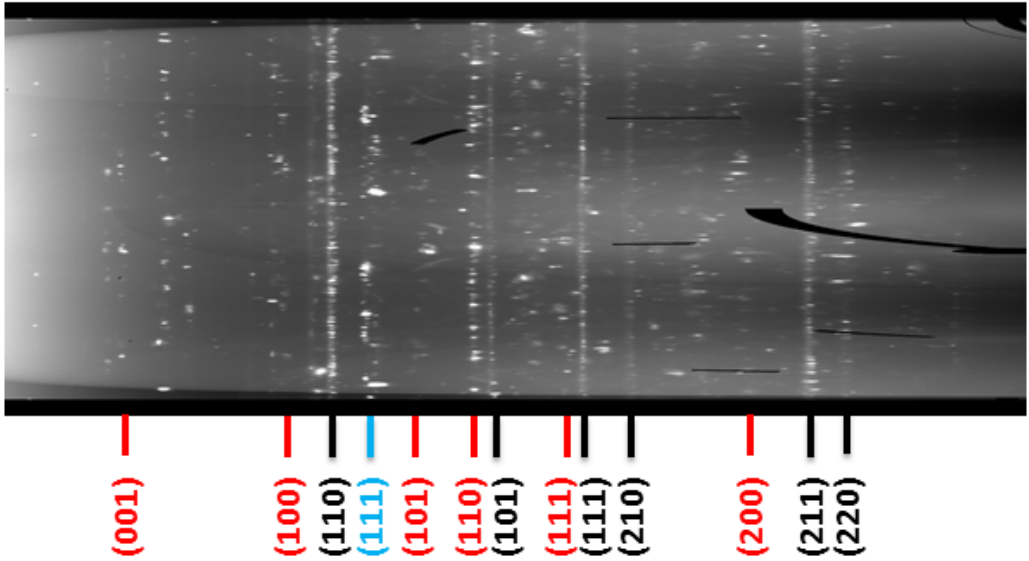

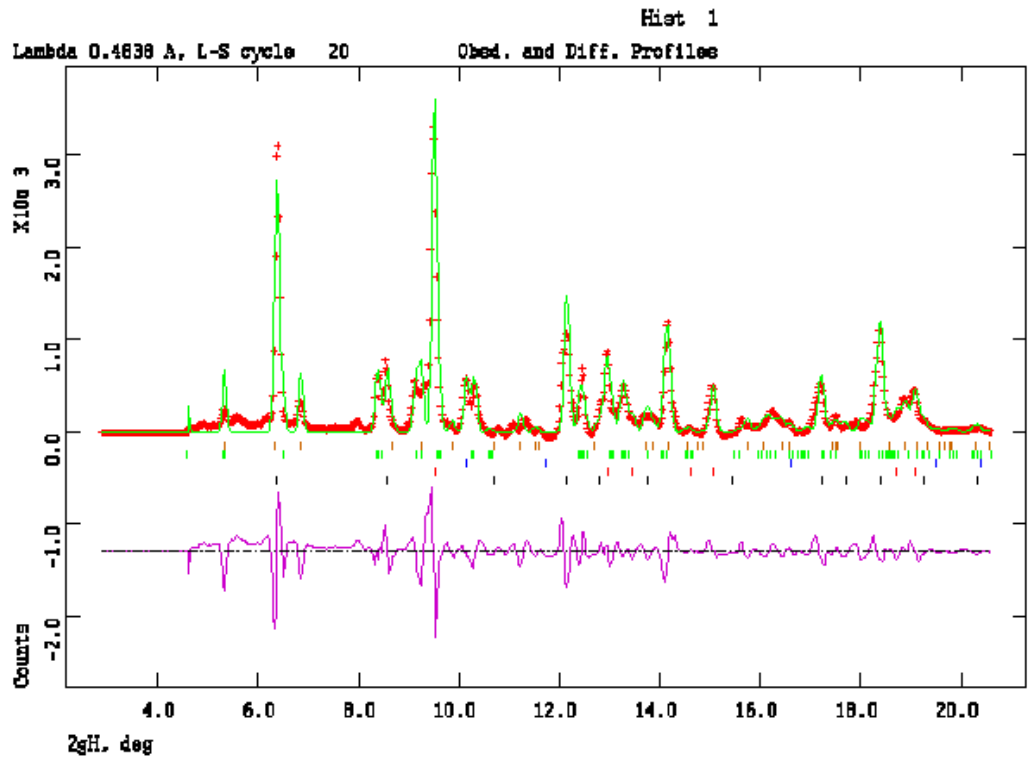

Fig. 2

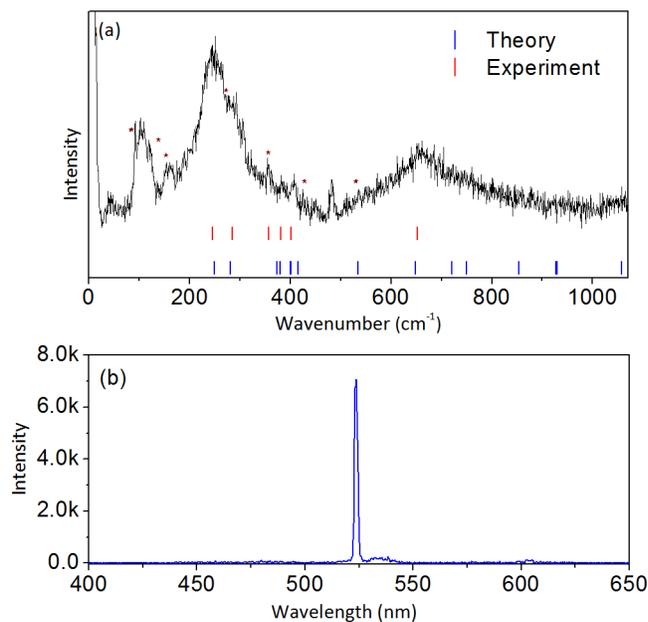

Fig. 3